\documentclass{jetpl}
\twocolumn

\lat

\title{Reconstruction of the Fermi surface in the pseudogap state of cuprates}

\rtitle{Reconstruction of the Fermi surface}

\sodtitle{Reconstruction of the Fermi surface in the pseudogap state of
cuprates}

\author{E.\,Z.\,Kuchinskii, M.\,V.\,Sadovskii\/\thanks{e-mail: sadovski@iep.uran.ru}}

\rauthor{E.\,Z.\,Kuchinskii, M.\,V.\,Sadovskii}

\sodauthor{Kuchinskii, Sadovskii }

\address{Institute for Electrophysics, RAS Ural Branch,
620016 Ekaterinburg, Russia}

\dates{*}{*}

\abstract{
Reconstruction of the Fermi surface of high-temperature superconducting
cuprates in the pseudogap state is analyzed within nearly exactly solvable model
of the pseudogap state, induced by short-range order fluctuations of 
antiferromagnetic (AFM, spin density wave (SDW), or similar charge density 
wave (CDW)) order parameter, competing with superconductivity.  We explicitly 
demonstrate the evolution from  ``Fermi arcs''  (on the ``large'' Fermi 
surface) observed in ARPES experiments at relatively high temperatures (when 
both the amplitude and phase of density waves fluctuate randomly) towards 
formation of typical ``small'' electron and hole ``pockets'', which are 
apparently observed in de Haas - van Alfen and Hall resistance oscillation 
experiments at low temperatures (when only the phase of density waves 
fluctuate, and correlation length of the short-range order is large enough).  
A qualitative criterion for quantum oscillations in high magnetic fields to 
be observable in the pseudogap state is formulated in terms of cyclotron 
frequency, correlation length of fluctuations and Fermi velocity.

}


\begin{document}

\maketitle

Pseudogap state of underdoped copper oxides 
\cite{Tim,MS,Tam,Nor07} is probably the main anomaly of the normal state of
high temperature superconductors. Especially striking is the observation of
``Fermi arcs'' in ARPES experiments, i.e. parts on the ``large'' Fermi surface 
around the diagonal of the Brillouin zone (BZ) with more or less well 
defined quasiparticles, while the parts of the Fermi surface close to BZ boundaries are 
almost completely ``destroyed'' \cite{Nor98, Shen03, Cam04}.  

However, the recent observation of quantum oscillation effects in
Hall resistance \cite{Tai1}, Shubnikov - de Haas \cite{Tai2} and de Haas - 
van Alfen (dHvA) oscillations \cite{Tai2,Tai3} in the underdoped YBCO 
cuprates, producing evidence for rather ``small'' hole or electron 
\cite{Tai4} pockets of the Fermi surface, seemed to contradict the well established ARPES 
data on the Fermi surface of cuprates.  

Qualitatvive explanation of this apparent contradiction was given in
Ref. \cite{Sing} within very simplified model of hole-like Fermi surface evolution under 
the effect of short-range AFM fluctuations. Here we present an exactly 
solvable model of such an evolution, which is able to describe continuous 
transformation of ``large'' ARPES Fermi surface with typical ``Fermi arcs'' at 
high-enough temperatures into a collection of ``small'' hole-like and 
electron-like ``pockets'', which form due to electron interaction with 
fluctuations of SDW (CDW) short-range order at low temperatures (in the
absence of any kind of AFM (or charge) long-range order). We also formulate
a qualitative criterion for observability of quantum oscillation effects in
high-magnetic field in this, rather unusual, situation.

We believe that the preferable ``scenario'' for pseudogap formation 
can be most likely based on the picture of strong scattering of the charge
carriers by short--ranged antiferromagnetic (AFM, SDW) spin fluctuations
\cite{MS,Tam}, i.e. fluctuations of the order parameter competing with
superconductivity. In momentum representation this scattering transfers 
momenta of the order of ${\bf Q}=(\frac{\pi}{a},\frac{\pi}{a})$ 
($a$ -- lattice constant of two dimensional lattice). 
This leads to the formation of structures in the one-particle spectrum, 
which are precursors of the changes in the spectra due
to long--range AFM order (period doubling).
As a result we obtain non--Fermi liquid like behavior of the spectral density 
in the vicinity of the so called ``hot spots'' on the
Fermi surface, appearing at intersections of the Fermi surface 
with antiferromagnetic Brillouin zone boundary \cite{MS,Tam}, which in the
low temperarure (large correlation length of the short-range order) can lead
to a significant Fermi surface reconstruction, similar to that appearing in
the case of AFM long-range order. 

Within this approach we have already demonstrated \cite{jetpl,Z_fac} the 
formation of ``Fermi arcs'' at high-enough temperatures, when AFM 
fluctuations can be effectively considered as static and Gaussian 
\cite{KS,Pin}. Here we present an exactly solvable model, quite similar to that
analyzed qualitatively in Ref. \cite{Sing}, which is capable to describe a
crossover from ``Fermi arc'' picture at high temperatures (typical for most
of ARPES experiments) to that of small  ``pockets'' at low temperatures
(typical for quantum oscillation experiments).

We shall consider a two - dimensional generalization of an exactly solvable
model proposed in one - dimension in Ref. \cite{BK} (and also analyzed in
a simplified two-dimensional approach in Ref. \cite{KSBK}), which is 
physically equivalent to the model of Ref. \cite{Sing}, but can produce a
complete picture of Fermi surface reconstruction and formation of both hole and electron
``pockets''.

We consider electrons in two-dimensional square lattice with nearest ($t$) and
next nearest ($t'$) neighbour hopping integrals, which leads to the usual
``bare'' dispersion: 
\begin{equation}
\varepsilon({\bf k})=-2t(\cos k_xa+\cos k_ya)-4t'\cos k_xa\cos k_ya-\mu\;\;,
\label{spectr}
\end{equation}
where $a$ is the lattice constant, $\mu$ --- chemical potential, and assume that these electrons are
scattered by the following (static) random field, imitating AFM(SDW) 
(or similar CDW) short-range order:  
\begin{equation}
V({\bf l})=D\exp(i{\bf Ql}-i{\bf ql})+D^*\exp(-i{\bf Ql}+i{\bf ql})
\label{Vx}
\end{equation}
where ${\bf l}=(n_xa,n_ya)$ numerates lattice sites and
$D = |D|e^{i\phi}$ denotes the complex 
amplitude of fluctuating SDW (or CDW) order parameter, while ${\bf q}=(q_x,q_y)$ 
is a small deviation from the dominating scattering 
vector ${\bf Q}=(Q_x,Q_y)=(\frac{\pi}{a},\frac{\pi}{a})$.

Generalizing the approach of Refs. \cite{BK,KSBK} (compare with Ref. 
\cite{Sing}) we consider a specific model of disorder, where both $q_x$ and 
$q_y$ are random and distributed according to:  
\begin{equation} 
{\cal P}(q_x,q_y)=\frac{1}{\pi^2}\frac{\kappa}{q^2_x+\kappa^2} 
\frac{\kappa}{q^2_y+\kappa^2}
\label{Lorxy}
\end{equation}
where $\kappa=\xi^{-1}$ is determined by the inverse correlation length of
short-range order. Phase $\phi$ is also considred to be random and distributed
uniformly on the interval $[0,2\pi]$.

Factorized form of (\ref{Lorxy}) is not very important 
physically, but allows for an analytic solution
for the Green' function which takes the form \cite{KSBK}:
\begin{equation} 
G_D(\varepsilon ,{\bf k})=\frac{\varepsilon -\varepsilon ({\bf k}+{\bf Q})+iv\kappa}
{(\varepsilon -\varepsilon ({\bf k}))(\varepsilon -\varepsilon ({\bf k}+{\bf Q})+iv\kappa )
-|D|^2}
\label{GrFun}
\end{equation}
where $v=|v_x({\bf k}+{\bf Q})|+|v_y({\bf k}+{\bf Q})|$, 
with $v_{x,y}({\bf k})=\frac{\partial\varepsilon ({\bf k})}{\partial k_{x,y}}$.

\begin{figure}
\includegraphics[clip=true,width=0.5\textwidth]{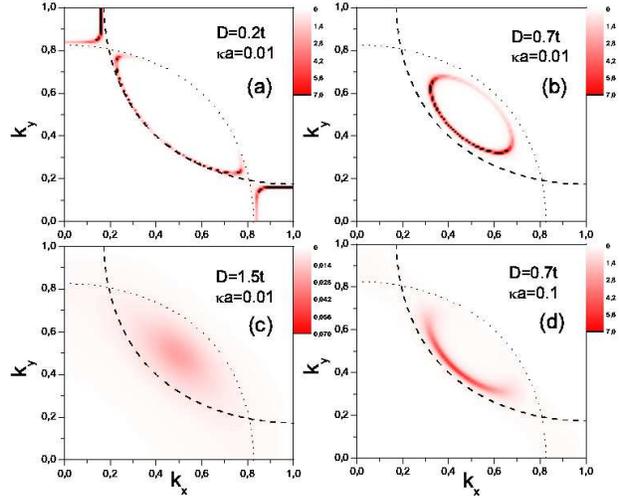}
\caption{Fig.1. Reconstruction of the Fermi surface in the low temperature
(large correlation length) regime of pseudogap fluctuations 
($n=0.9$, $t'/t=-0.4$. Shown are intensity plots of spectral 
density for $\varepsilon=0$:
(a) -- $D=0.2t$, $\kappa a=0.01$;\ (b) -- $D=0.7t$, $\kappa a=0.01$;\ 
(c) -- $D=1.5t$, $\kappa a=0.01$;\ (d) -- $D=0.7t$, $\kappa a=0.1$;\ 
Dashed line denotes ``bare'' Fermi surface, dotted line --- shadow Fermi 
surface.}
\label{Reconstruct}
\end{figure}

Spectral density 
$A(\varepsilon ,{\bf k})=-\frac{1}{\pi}ImG_D(\varepsilon ,{\bf k})$
at the Fermi level ($\varepsilon =0$), is shown in Fig. \ref{Reconstruct}, 
and demonstrate the formation of small ``pockets'' instead of large ``bare'' 
Fermi surface. Here and in the following we have assumed rather typical (for
cuprates) values of $t'/t=-0.4$ and doping $n=0.9$ (10\% hole doping), 
corresponding to $\mu =-1.08t$.

The poles of the Green's function (\ref{GrFun}), determining the quasiparticle
dispersion and damping the limit of large enough correlation length 
($v\kappa\ll t$, low temperature), are given by:
\begin{equation}
\tilde E^{(\pm)}=E^{(\pm)}_{\bf k}-i\frac{v\kappa}{2}
\left( 1\mp\frac{\varepsilon^{(-)}_{\bf k}}{E_{\bf k}}\right)
\label{poleGF}
\end{equation}
with $\varepsilon^{(\pm)}_{\bf k}=\frac{1}{2}[\varepsilon ({\bf k})\pm 
\varepsilon ({\bf k}+{\bf Q})]$,\  
$E_{\bf k}=\sqrt{\varepsilon^{(-)2}_{\bf k}+|D|^2}$, and
\begin{equation} 
E^{(\pm)}_{\bf k}=\varepsilon^{(+)}_{\bf k}\pm \sqrt{\varepsilon^{(-)2}_{\bf k}+|D|^2}
\label{EnergyDisp}
\end{equation}
which is just the same as dispersion in the case of the presence of long-range AFM order.
Equation $E^{(-)}_{\bf k}=0$ determines the hole ``pocket'' of the Fermi
surface, around the point $(\frac{\pi}{2a},\frac{\pi}{2a})$ in the Brillouin
zone, while $E^{(+)}_{\bf k}=0$ defines the electronic ``pockets'', centered 
around $(\frac{\pi}{a},0)$ and $(0,\frac{\pi}{a})$, as shown in 
Fig. \ref{Reconstruct} (a).

Quasiparticle damping as given by the imaginary part of (\ref{poleGF}) 
is, in fact, changing rather drastically as particle moves around the ``pocket'' 
of the Fermi surface. Being practically zero in the nearest to point
$\Gamma=(0,0)$  nodal (i.e. on the diagonal of the Brillouin zone) point of this
trajectory on  the hole ``pocket'', it becomes of the order of
$\approx v_F^n\kappa$ in the far (from $\Gamma$) nodal point. Here we have
introduced
$v_F^n=\left.|v_x({\bf k}|)+
|v_y({\bf k})|\right|_{\varepsilon ({\bf k})=0, k_x=k_y}$ 
--- particle velocity at the nodal point of the ``bare'' Fermi surface. 
On the trajectory around the electronic ``pocket'' quasiparticle damping
changes from nearly zero near the crossing points  of the ``bare'' Fermi surface
%($k_x=\frac{\pi}{a}$ or $k_y=\frac{\pi}{a}$) 
with Brillouin zone boundary up to $\approx v_F^a\kappa$ at points 
%$k_x=0$ or $k_y=0$, 
close to the similar crossing points of the ``shadow'' Fermi surface. 
Here $v_F^a=\left.|v_x({\bf k}|)+|v_y({\bf k})|\right|
_{\varepsilon ({\bf k})=0, k_x=\frac{\pi}{a}}$ is the velocity in the antinodal 
point of ``bare'' Fermi surface.

Of course, the complete theory of quantum (Shubnikov - de Haas or
de Haas - van Alfen) oscillations for such peculiar
situation can be rather complicated. However, a rough
qualitative criterion for the observability of quantum oscillations in our model 
can be easily formulated as follows. 
Effective width of spectral densities in our model, which determines 
smearing of the Fermi surfaces, can be roughly compared to 
impurity scattering contribution to Dingle temperature and estimated as 
$\tau^{-1}\sim \frac{<v_F>}{\xi}$, where $<v_F>$ is the velocity averaged over the Fermi 
surface. In fact it gives a kind of the upper boundary to pseudogap scattering 
rate in our model. Then our criterion takes the obvious form:
\begin{equation}
\omega_H\frac{\xi}{<v_F>}\sim\frac{\omega_H}{t}\frac{\xi}{a}\gg 1
\label{criterion}
\end{equation}
where $\omega_H$ is the usual cyclotron frequency.

As the most unfavourable estimate (overestimating the effective damping)
we take:
\begin{equation} 
<v_F>=\left\{ \begin{array}{lcr} 
v_F^n  \mbox{  for hole ``pocket'' } \\
v_F^a  \mbox{  for electronic ``pocket'' }
 \end{array}\right.
\label{estimatVf}
\end{equation}
Experimentally oscillations become observable in magnetic fields larger than
50 T \cite{Tai1,Tai2,Tai3,Tai4}. Taking the large correlation length
$\xi =100\, a$ and magnetic field $H=$50T we get $\omega_H\tau \approx 0.8$ 
for hole ``pocket'' and $\omega_H\tau \approx 1.3$ for electronic ``pockets''
in our model. Thus we need rather large values of correlation length
$\xi \sim 50-100\, a$ for oscillations to be observable. However, this value may
be smaller in the case of cyclotron mass larger than the mass of the free 
electron used in the above estimates. 

%Let us briefly discuss the problem of estimating the density of carriers from
%the frequency of quantum oscillations, i.e. from the area of 2d ``pockets''
%of the Fermi surface. 
From Luttinger theorem it follows that the number
of electrons per cell is given by $n=2a^2\frac{S_{fs}}{\pi ^2}$, 
where $S_{fs}$ is the area of the ``bare'' Fermi surface 
($\varepsilon ({\bf k})=0$) in the quarter of the Brillouin zone. 
Similarly, we can determine this concentration as 
$n=2a^2\frac{S_{sh}}{\pi ^2}$ calculating
the area $S_{sh}$ of the ``shadow'' Fermi surface 
($\varepsilon ({\bf k}+{\bf Q})=0$) around the point 
$M(\frac{\pi}{a},\frac{\pi}{a})$. Obviously $S_{sh}=S_{fs}$. 
Then, in the limit of $|D|\to 0$, for hole doping we get \cite{ChKee,Morin}:
\begin{equation} 
p=1-n=a^2\frac{S_h-S'_e}{\pi ^2}=a^2\frac{S_h-S_e/2}{\pi ^2}
\label{doping}
\end{equation}
where $S_h$ is the area of hole ``pocket'' and  $S'_e$ is the area of the parts
of electronic pocket inside the quarter of the Brillouin zone (which is a half
of the total area of electronic ``pocket'' $S_e$). 

\begin{figure}
\includegraphics[clip=true,width=0.4\textwidth]{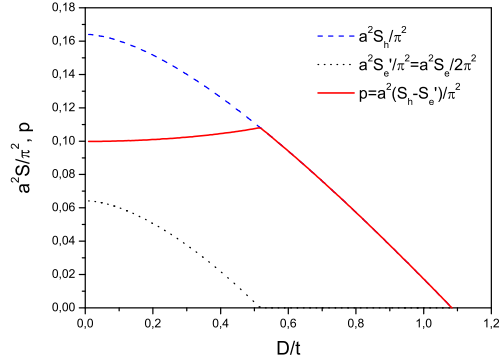}
\caption{Fig.2. The area of hole ($a^2\frac{S_h}{\pi ^2}$) and
electronic ($a^2\frac{S'_e}{\pi ^2}$) ``pockets'' in the quarter of Brillouin 
zone and ``doping'' $p=a^2\frac{(S_h-S'_e)}{\pi ^2}$ as fuctions of the
pseudogap amplitude $D/t$ ($n=0.9$ ($\mu =-1.08t$), $t'/t=-0.4$).}
\label{figdoping}
\end{figure}

However, these expressions are valid only for $|D|\to 0$. With the growth of the
pseudogap amplitude $|D|$ the area of both hole and electronic  
``pockets'' diminish (as can be seen from Fig.\ref{Reconstruct} and 
Fig.\ref{figdoping}). In the presence of electronic ``pocket'' this suppression
of the area of both ``pockets'' compensate each other, leaving the doping given by
Eq. (\ref{doping}) almost unchanged (Fig.\ref{figdoping}). After the
disappearance of electronic  ``pocket'', taking place at $|D|=\mu -4t'=0.52t$ 
(i.e. when $E^{(+)}_{{\bf k}=(\pi /a,0)}=0$), there is no way to compensate
the suppression of the area the hole ``pockets'' with the growth of  
$|D|$ and the number of carriers, determined by (\ref{doping}), will also be
suppressed, going to zero with the disappearance of the hole ``pocket'',
taking place at $|D|=-\mu =1.08t$ 
(which is defined by $E^{(-)}_{{\bf k}=(\pi /2a,\pi /2a)}=0$) and dielectric
(AFM) gap ``closes'' the whole Fermi surface (Fig.\ref{Reconstruct}(c)). 
Thus, the doping calculated according to Eq. (\ref{doping}) in the case of large 
enough pseudogap amplitude (in the absence of electronic ``pocket'') will be
significantly underestimated.

Experimentally, only one frequency of quantum oscillations 
$F\approx 540 T$ was observed in YBCO \cite{Tai2}. Assuming it corresponds
to the presence of only the hole ``pocket'', we obtain for the
area of this ``pocket''  $a^2S_h/\pi ^2=0.078$, which, according to
Fig.\ref{figdoping} corresponds to $|D|\approx 0.7t$.

%However, we can not exclude the possibility that in the experiments only
%contribution from electron ``pockets'' is observable due to suppression of
%hole ``pocket'' contribution due to breaking of our criterion for the
%observability of quantum oscillations. 

\begin{figure}
\includegraphics[clip=true,width=0.5\textwidth]{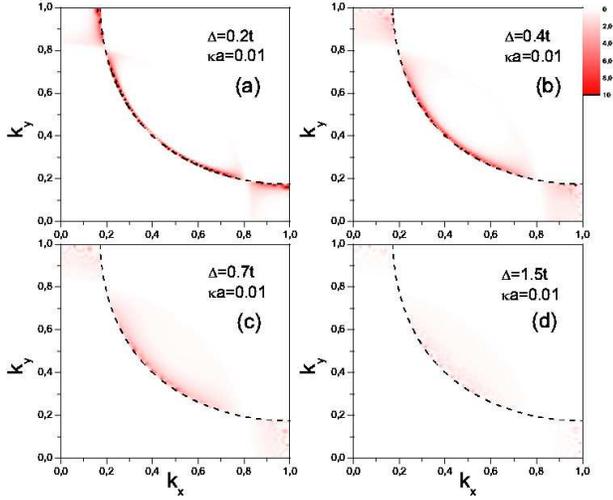}
\caption{Fig.3. Formation of the Fermi ``arcs'' in the high-temperature
regime of  pseudogap  fluctuations ($n=0.9$, $t'/t=-0.4$, $\kappa a=0.01$).
Shown are intensity plots of spectral density for 
$\varepsilon=0$. 
(a) -- $\Delta=0.2t$;\ (b) -- $\Delta=0.4t$;\ (c) -- $\Delta=0.7t$;\ (d) -- $\Delta=1.5t$;\ 
Dashed line denotes ``bare'' Fermi surface.}
\label{Destruct}
\end{figure}

Green's function (\ref{GrFun}) describes the ``low temperature'' regime of
pseudogap fluctuations, when the amplitude fluctuations of the random field
(\ref{Vx}) are ``frozen out''. In the ``high temperature'' regime both the phase
and the amplitude $|D|$ of (\ref{Vx}) are fluctuating. Assuming these fluctuations 
Gaussian we take the probability distribution of amplitude fluctuations given
by Rayleigh distribution \cite{KSBK}:
\begin{equation} 
{\cal P}_D(|D|)=\frac{2|D|}{\Delta ^2}exp\left( -\frac{|D|^2}{\Delta ^2}\right)
\label{Relay}
\end{equation}
Then the averaged Green's function takes the form:
\begin{equation} 
G_{\Delta}(\varepsilon ,{\bf k})=\int_{0}^{\infty}d|D|{\cal P}_D(|D|)
G_D(\varepsilon ,{\bf k})
%\frac{\varepsilon -\varepsilon ({\bf k}+{\bf Q})+iv\kappa}
%{(\varepsilon -\varepsilon ({\bf k}))(\varepsilon -\varepsilon ({\bf k}+{\bf Q})+iv\kappa )
%-|D|^2}
\label{GrFun1}
\end{equation}
Profiles of the spectral density at the Fermi level ($\varepsilon =0$),
corresponding to (\ref{GrFun1}) and different values of the pseudogap width
$\Delta$ are shown in Fig. \ref{Destruct}. The growth of the pseudogap width
leads to the ``destruction'' of the Fermi surface close to Brillouin zone
boundaries and formation of typical Fermi ``arcs'', qualitatively (and
quantitatively) similar to that obtained in our previous work 
\cite{jetpl,Z_fac} and in accordance with the results of ARPES experiments, which are typically 
done at much higher temperatures, than experiments on quantum fluctuations.

This work is supported by RFBR grant 08-02-00021 and RAS programs 
``Quantum macrophysics'' and ``Strongly correlated electrons in 
semiconductors, metals, superconductors and magnetic materials''. MVS
is gratefully acknowledges a discussion with L. Taillefer at GRC'07 Conference
on Superconductivity, which stimulated his interest in this problem.

\pagestyle{empty}

\end{document}